\documentclass[10.5pt]{article}
\usepackage[margin=0.65in]{geometry}
\usepackage[utf8]{inputenc}
\usepackage{graphicx}
\usepackage{authblk}
\usepackage{caption}
\usepackage{subcaption}

\title{What's for Lunch? \\ \Large A systematic ordering of foods in the Soup-Salad-Sandwich phase space}
\author[1]{Madelyn Leembruggen}
\author[2]{Caroline Martin}
\affil[1]{\small \textit{Department of Physics, Harvard University, Cambridge, MA 02138, USA}}
\affil[2]{\small \textit{Harvard John A. Paulson School of Engineering and Applied Sciences, Harvard University, Cambridge,
MA 02138, USA}}
\date{April Fools' Day 2022}

\begin{document}

\maketitle

\begin{abstract}
The statistical physics of phase transitions has been hugely successful at describing numerous natural, physical, and technological phenomena, and now a rigorous examination of the phase space of the culinary regime is likewise ripe for the picking. Despite great demand for the resolution of many scientific debates over the taxonomy of food, past attempts have failed to account for complex phase behavior and co-existence, and have thus left the public hungry for a more substantial theory. By applying the principles of statistical physics and thermodynamics, we here map out the complete phase space of all culinary dishes and find three distinct phase regimes: Soup, Salad, and Sandwich. We consider the effect of different state variables on these phase boundaries, as well as regions of co-existence and triple points. With this complete 3-dimensional phase diagram of all foods, we can conclusively answer many bitter debates, including the imperishable question ``is a hotdog a sandwich?'' The answer: yes.
\end{abstract}

\section{Introduction} 
\label{intro}
Phase transitions are ubiquitous across the physical world. By understanding the underlying physics of phases, we have been able to make connections between such disparate behavior as the ordering of spins in a lattice \cite{cardy_scaling_1996}, the melting of sea ice \cite{golden_thermal_2007}, and the firing of neurons in a brain \cite{tkacik_thermodynamics_2015}. The similar behavior of such different systems which share a single universality class enables not only physical comparison, but also increased predictive power for such asymptotic phenomena as the critical exponents. 

In recent years, the tools developed in the study of the phase behavior of physical systems have been zestfully applied to food science, though unfortunately with great controversy and dissent. Namely, the as-yet unresolved scientific debate of “Is a hotdog a sandwich?” has periodically sparked a resurgence of interest in this field. However, all efforts to this point have floundered and fallen short of a complete examination of the full phase behavior of foods, neglecting such important concepts as phase transitions, co-existence, and critical points within the phase space. Without this statistical physical framework, there have been grave misunderstandings and misclassifications. In fact, many fierce disputes over the phase of a particular dish can be solved simply by consider regions of phase co-existence. 

Here, we fully map out the phase space of the soup-salad-sandwich (Triple S) transitions along the axes that we have identified as essential for studying 
these three phases: effective temperature, effective pressure, and carbohydrate enclosure. We then identify order parameters, including effective density of the matrix and carbohydrate ordering. 

Some might wonder, ``why address this question at all?'' In answer, we point to the historic scientific progress and advancements made in the fields of thermodynamics and statistical physics. Innumerable interesting properties of systems have been explored and examined through the lens of phase transitions and critical phenomena. The same progress can also be made in the vast menu of culinary phenomena. By identifying distinct phases, we reveal common properties for each phase, and although some of these characteristics might be abstract, practical conclusions can also be drawn. For example, knowing the phase of a dish may allow you to adequately prepare for your meal by predicting the types of utensils which will be necessary, or the potential for the dish as a finger food. More excitingly, studying the transitions between phases may lead to bizarre and novel dishes which are incapable of existing anywhere except at the phase boundaries.

Finally, we are motivated to carry out this analysis to achieve a more universal understanding of dishes, one that rises above the arbitrary limitations of human language. Often, different cultures will categorize fundamentally similar foods in vastly disparate ways. While some types of cuisine might not formally have ``sandwiches'', they certainly have foods that could, in a broader sense, be classified as such. Thus, although we frame our problem in terms of American-centric definitions of “soup”, “salad”, and “sandwich”, we aim to abstract away our colloquial connotations and provide objective and quantitative definitions useful for characterizing these phases and their critical behavior.

\begin{figure}[h]
    \centering
    \includegraphics[width=0.6\textwidth]{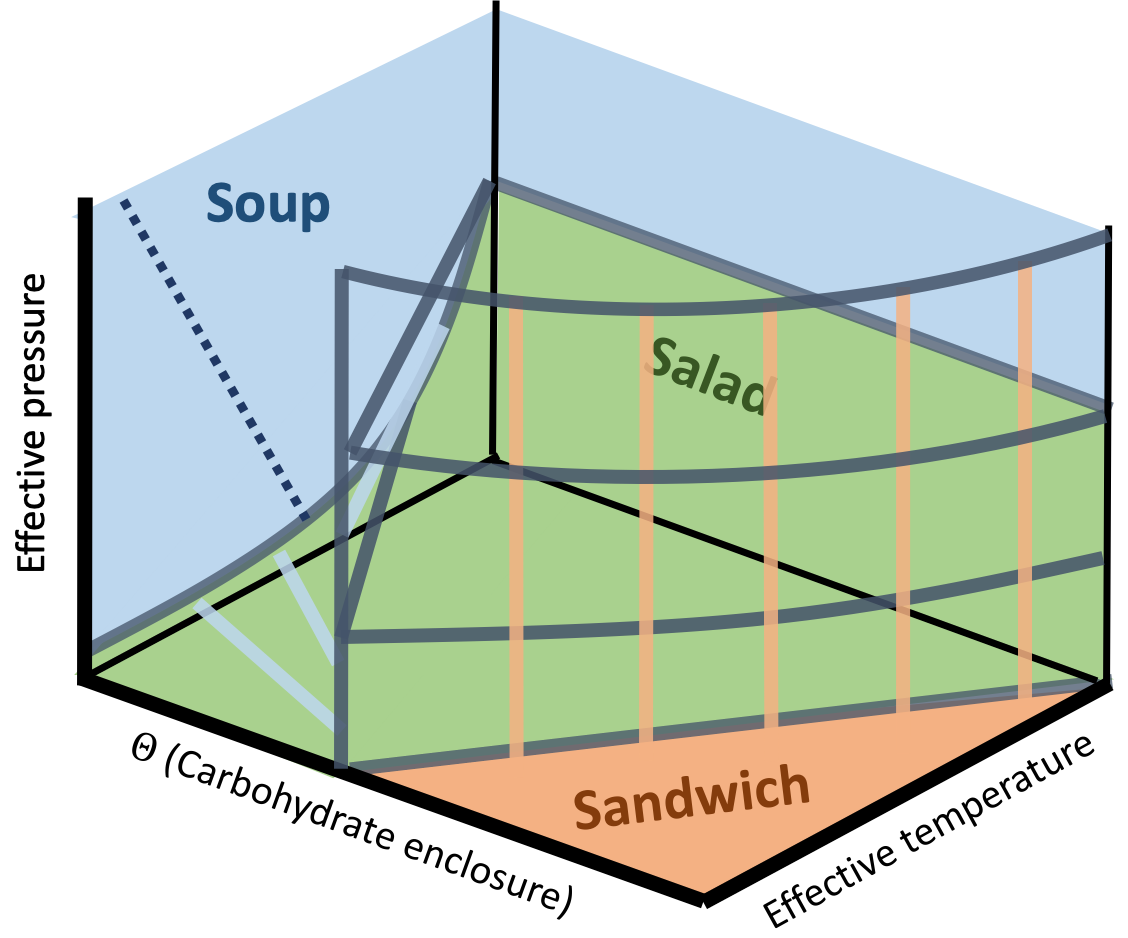}
    \caption{The full phase space of all foods can be broken down into three distinct regions of phase space: soup, salad, and sandwich. These phase regions depend on such physical parameters as effective temperature, effective pressure, and carbohydrate enclosure.}
    \label{fig:phase_diagram}
\end{figure}

\subsection{Background}
\begin{figure}[h]
    \centering
    \includegraphics[width=\textwidth]{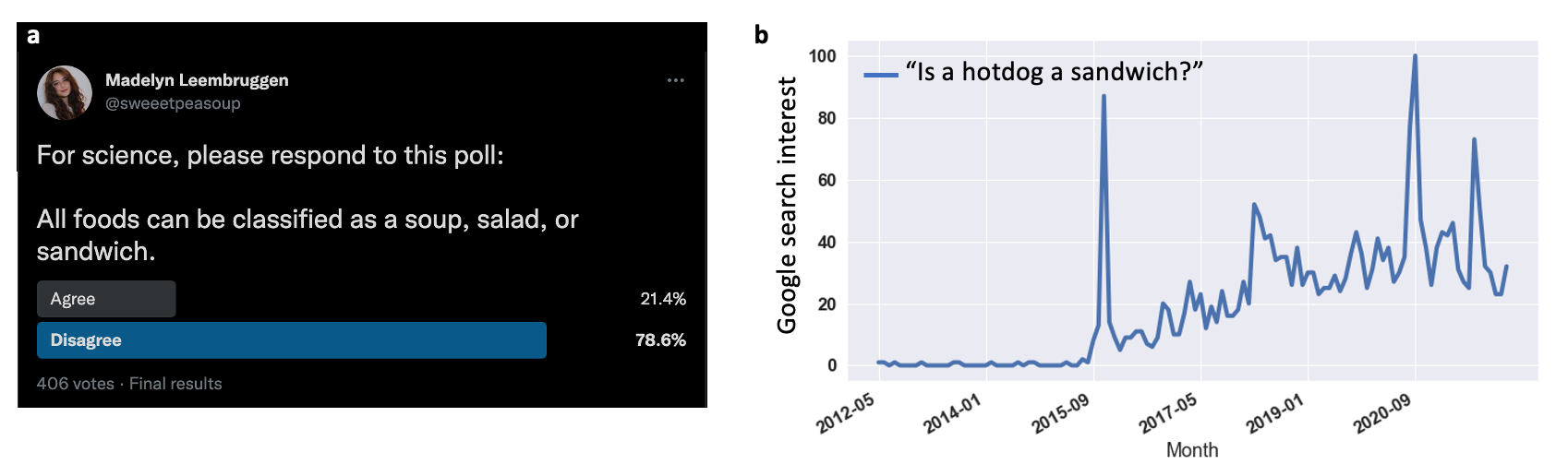}
    \caption{Attempts of culinary taxonomy and phase categorization have generated widespread public interest, though with limited success. (a) A Twitter poll reveals broad distrust in the Triple S trichotomy, likely due to misguided past attempts at applying these descriptors without a full understanding of the Triple S phase behavior. (b) Many of the previous research attempts can be traced back to an effort to resolve The Great Hotdog-Sandwich Debate: ``Is a hotdog a sandwich?''. This issue periodically recaptures the zeitgeist, indicating strong public appetite for a conclusive answer.}
    \label{fig:twitter_poll}
\end{figure}

In many ways, the viral popularity of The Great Hotdog-Sandwich Debate revolutionized the field of culinary taxonomy. Because it has never been fully settled, interest in this question periodically re-emerges in the zeitgeist, as demonstrated in Figure \ref{fig:twitter_poll} (b). Previous attempts to formalize an answer to this question have made some progress despite various flaws in logic and scientific understanding. One important intellectual jump was the realization that food exists in more categories than just sandwich alone. Broader categories were introduced, such as the Soup-Salad-Sandwich (Triple S) trichotomy \cite{noauthor_soup_nodate} and the Cube Rule \cite{noauthor_cube_nodate}, which included such categories as Toast, Sandwich, and Taco. Others have introduced the concept of foods existing in regions of many dimensions, included proposed dimensions such as Soupiness, Arrangement Entropy, and Ingredient Entropy \cite{noauthor_salad_nodate}. While we acknowledge the intellectual progress made in each of these attempts, we feel that none have satisfyingly resolved the debate; of them, however, the authors (along with many other citizen scientists participating in social media discourse) are most compelled by the attempt to classify \textit{all} foods as either a soup, salad, or sandwich.

Despite our ostensibly innate human desire to sort and classify information, past efforts to apply this Triple S trichotomy to the culinary space have continued to lose steam. This has resulted in widespread skepticism in the validity of the approach at all. In a Twitter poll conducted over a 24 hour period with 406 responses, only $21.4\%$ of our respondents agreed that all foods can be classified as a soup, salad, or sandwich (Fig. \ref{fig:twitter_poll}). Most of the remaining $70.6\%$ of respondents rather jarringly disagreed. For example, Twitter user @I\_Am\_Stan, who revealed they strongly disagreed, stated ``Please vote in this nonsense [poll]. There is only one correct choice.'' Amongst even the more mild dissenters, distaste was immediate, as reported by Twitter user @BeckyDouglas: ``I have clicked `disagree' but I want you to know that this has sparked a heated debate in the household this evening.''

Our public opinion poll (Figure \ref{fig:twitter_poll} (a)) reveals a strong distrust of the Triple S trichotomy. We draw the conclusion that, as a trichotomy, soup-salad-sandwich is emphatically unfavorable; attempts to sort foods into three distinct, non-overlapping bins are controversial and fail time and again. However, these authors believe that all hope is not lost. We assert that public discourse has been focusing on the wrong question. The question should not be ``can all foods be \textit{sorted}'', but rather ``can all foods be \textit{classified}''. Although subtle, the distinction of \textit{classification} unfurls an entire phase space for foods where soup, salad, and sandwich are phases separated by first or second order transitions, with numerous regions of co-existence and criticality. It is in these transitional regions that the most wonderful (and sometimes the most appalling) dishes reside.

Other attempts to resolve this question of food classification have included mapping all dishes onto a 3-dimensional space where each dish has its own values for the soupiness, sandwichness, and saladness bounded between 0 and 1 \cite{noauthor_soup-salad-sandwich_nodate}. While interesting, this effort misunderstands the role here of Soup, Salad, and Sandwich. They are not \textit{state variables}, they are distinct \textit{phases}. Because this previous work demonstrates a fundamental misunderstanding of phase behavior, the result is a vast oversimplification of the Triple S space. 

\subsection{Methods}
In this paper we take a different approach, and fully map the Triple S space and phase transitions along three axes identified as essential for capturing the appropriate phenomena of the space: effective temperature ($\tau$), effective pressure ($\rho$), and carbohydrate enclosure ($\Theta$). By varying these state variables, we can move into different regions of the phase space. Importantly, in this work we only classify complete dishes-- a dish defined colloquially as something one would actually eat and prepared as such, not just a random individual ingredient. This is because a single ingredient of a dish\footnote{Twitter user @gooseus replied to our poll (Fig. \ref{fig:twitter_poll}) with the question, ``What about a single berry or steak alone? Is that a salad of one item? a sandwich of organic tissues? or sandwich, holding all but one component?'', which we assert are single ingredients that cannot hold a phase on their own.} should be considered as analogous to an individual spin. Individual spins do not have a phase; rather, you must consider them as a collective group in order to characterize their bulk behavior. Therefore you cannot qualify the phase of single spin systems (i.e. single ingredient dishes). The individual component must be coupled to other components in order for the system (dish) as the whole to have a phase.

\section{Soup-Salad phase diagram} \label{soup_salad}

First, we consider a projection of the phase space which involves only the soup and salad phases. This is equivalent to a projection along the $\Theta = 0$ plane, with effective temperature and pressure ($\tau$ and $\rho$) as the remaining axes. We find that in this limit the soup-salad phase behavior maps almost directly to the phase diagram for water.

The order parameter for the soup-salad transition is the density of the embedding matrix. More explicitly, soups are a suspension of ingredients in a liquid matrix, whereas salads are a suspension of ingredients in a gaseous matrix. Thus, the soup-salad phase transition looks quite similar to the liquid-gas transition for water, with the primary difference that the soup-salad transition is a line of second order phase transitions for all $\tau$, (i.e. the critical temperature is $\tau_c = 0$).

Intuition for this distinction of soup from salad can be gained by considering an argument about pore size and penetration depth of the embedding matrix. If the bulkiest components of the dish form a scaffold, then the ability of the remaining ingredients to efficiently permeate this scaffold determines the phase of the dish. That is, if the constituents are completely incapable of penetrating the bulky scaffold, the dish must be a salad because the embedding matrix is considered to be air. On the other hand, complete penetration of the scaffold with another of the dish's ingredients must be a soup, because the embedding matrix is liquid in nature and thus able to permeate the system without flow restrictions.

Importantly, we identify the region of the soup-salad phase that is analogous to solid water as a \textit{glassy} soup. In the case of a glassy soup, the embedding matrix is not air and yet the dish, at its serving temperature, does not flow like a (liquid) soup. Glassy soups are separated from the soup phase by a line of first order phase transitions as the embedding matrix melts from a glass to a liquid. However, it should be noted that soups and glassy soups share an overwhelming number of properties and should not be treated as separate phases.

\begin{figure}[h]
    \centering
    \includegraphics[width=.9\textwidth]{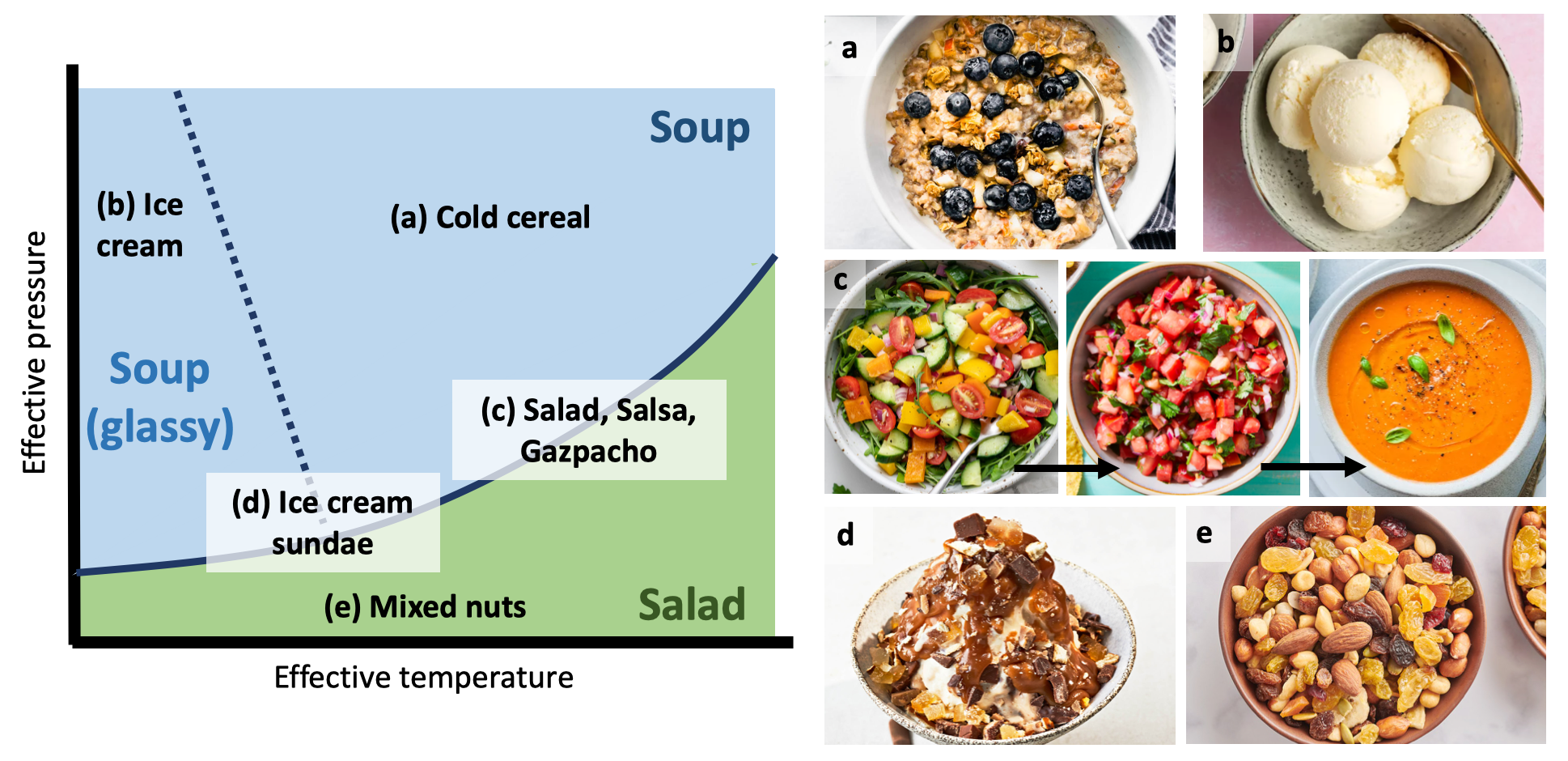}
    \caption{The Soup-Salad projection of the phase space (setting carbohydrate enclosure $\Theta = 0$) maps almost directly to the phase diagram for water, where the density of the dish's embedding matrix determines the phase of the dish. A liquid matrix indicates a soup (a), here shown as a bowl of cereal. A soup at low effective temperatures (frozen or solid embedding matrix) is classified as a \textit{glassy} soup (b), here shown as vanilla ice cream. If the components of a dish are embedded in a gaseous matrix, they are in the salad phase. The transition between the soup and salad phases is continuous, as demonstrated in (c) where an identical set of ingredients is presented as a chunky salad, a salsa, then finally as a gazpacho, smoothly transitioning from one phase to the other as the components of the dish are chopped more finely. The triple point of this phase diagram is embodied in an ice cream sundae (d), a dish in which all three phases coexist. Finally a salad is defined as a collection of components which are embedded in air, here shown as a bowl of mixed nuts (e).}
    \label{fig:soup-salad}
\end{figure}

\subsection{Case Studies} \label{soup_salad_cases}

In addition to colloquially agreed upon soups, some other examples of soup include smoothies, cold cereal, and hot oatmeal. Many types of salads are already in the common vernacular, such as leafy green salad, fruit salad, and tuna salad. Our results do indicate that many unexpected dishes are in fact salads, including trail mix, a handful of M\&M's, a plate of chicken wings\footnote{Thank you to Twitter user @supernova\_mads for raising this question and to Twitter user @biguyreacts who suggested ``chicken skin as the crust so chicken wings are a sandwich'' but was unfortunately wrong, as will be explained in more detail in Section \ref{sandwich}.}, or even shredded cheese eaten straight from the bag while you stand in front of your open fridge at 2:30 AM. Glassy soups are actually quite prevalent; a few examples include ice cream, chocolate, and hummus\footnote{Thank you to Twitter user @QuantumYakar for asking about the phase of hummus.}. Note that some dishes colloquially referred to as salads (such as tuna, chicken, egg, and potato ``salad'') often exist close to the salad-glassy soup boundary and will experience a second order phase transition to a glassy soup as a function of the amount of mayonnaise added to the salad (note that mayonnaise on its own should be classified as a glassy soup). Similarly, a bowl of cubed Jello sits near the salad-glassy soup transition, as the Jello itself is a glassy soup, but the cubes in a bowl constitute a salad (as the embedding matrix is air). On the other hand, a Jello ``salad'' (which is comprised of fruits and vegetables embedded in a Jello matrix) is actually not a salad at all, but rather a glassy soup (though as the Jello matrix cools from a liquid into a gel, the Jello salad will cross the soup-glassy soup boundary). 

We now consider the second order phase transition which occurs across the soup-salad boundary. On one side of the transition, we find such dishes as a chunky salad with tomatoes, onions, and peppers with a dressing of lime juice and olive oil. If the same ingredients are diced more finely, the same dish might be called salsa, and salsas span a wide range from chunky to soupy. By blending the same ingredients you can make gazpacho, which falls solidly in the soup phase (Fig \ref{fig:soup-salad} (c)).

Finally, as an example of the triple point of the soup-salad phase space, we present the ice cream sundae. It is built from both a soup (hot fudge) and glassy soup (ice cream) with salad components (sprinkles or candy bits) and arranged as a salad overall. In this narrow co-existence region, we find this nonpareil of food, the pinnacle of culinary ingenuity, a shining example of the edible richness discovered through the insights of phase behavior.

\section{The Sandwich axis} 
\label{sandwich}

We now introduce a third axis and state variable: the carbohydrate enclosure ($\Theta$) of the system. $\Theta$ is bounded from $[0,1]$, with $0$ indicating that the carbohydrate is completely enclosed by the other elements of the dish (henceforth referred to as ``elements''), and $1$ indicating the state where all elements are contained entirely within the carbohydrate. Our carbohydrate enclosure parameter is similar in spirit to the Cube Rule, which was introduced and popularized in 2018 \cite{noauthor_cube_nodate}. The Cube Rule differentiates between sandwiches and not-sandwiches based on where the carb is oriented with respect to the other elements. While we draw a great amount of inspiration from the Cube Rule, we assert their definition of ``sandwich'' is in fact too narrow, and most of the food categories they identify actually exist in the sandwich phase. The following analysis will elucidate the breadth of the sandwich phase, and provide compelling arguments for carbohydrate enclosure as the proper parameter for this third axis.

\begin{figure}[h]
    \centering
    \includegraphics[width=\textwidth]{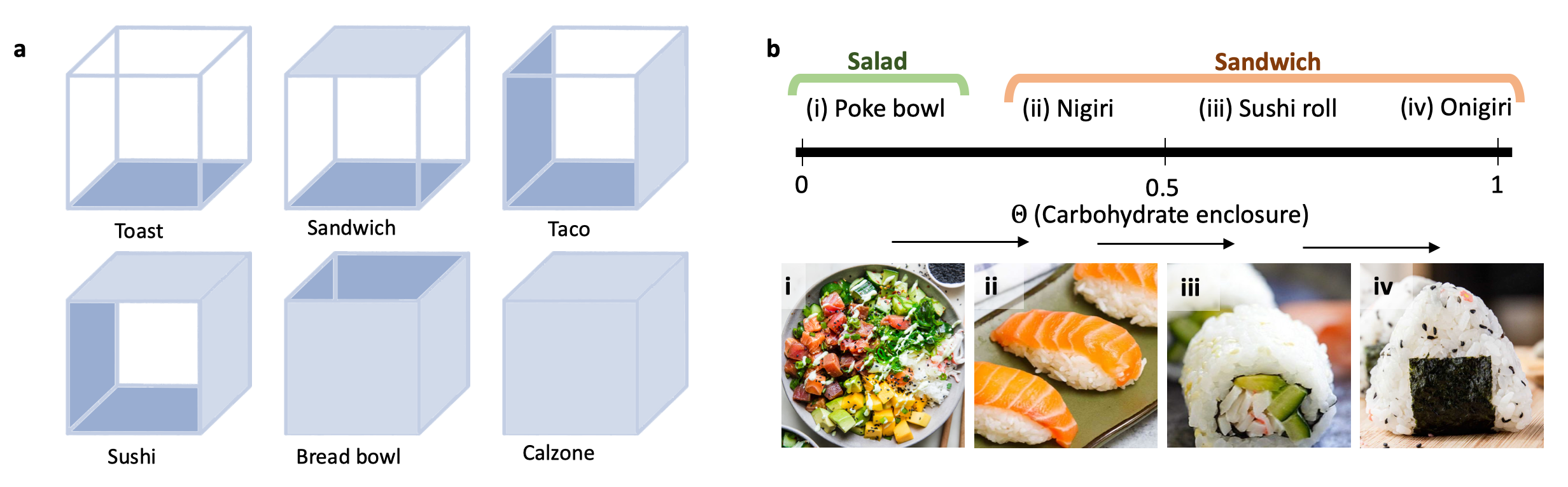}
    \caption{(a) We introduce the state variable of the carbohydrate enclosure $\Theta$, inspired by past work which considered the effect of carb enclosure in food taxonomy: the Cube Rule \cite{noauthor_cube_nodate}. We reject the differentiation of carb-containing foods into these six categories and feel it is an unnecessary complication of the phase space. (b) Instead, we introduce the parameter of the the carbohydrate enclosure ($\Theta$). Here, as an introduction to the role of $\Theta$ as a state variable, we consider the transition of the same ingredients from salad (poke bowl) to sandwich (nigiri, sushi roll, and onigiri) simply by varying the enclosure of carbohydrate (rice).}
    \label{fig:carb_enclosure}
\end{figure}

Interestingly, the sandwich region is not bounded by a constant carbohydrate enclosure. Instead, the location of this phase transition depends also on the effective pressure and temperature of the system. This is because the efficacy of a carbohydrate at enclosing the elements depends on the effective temperature and pressure of the elements. If the elements are in a soup phase (high effective pressure and temperature), a carbohydrate enclosure close to $1$ will be necessary for that dish to be in sandwich phase. In the more solid region of the TP plane (generally lower effective pressure and temperature), on the other hand, a dish can be in the sandwich phase with a relatively lower value of $\Theta$, such as an open-face sandwich. 

\begin{figure}[h!]
    \centering
    \begin{subfigure}{0.75\textwidth}
        \centering
        \includegraphics[width=\textwidth]{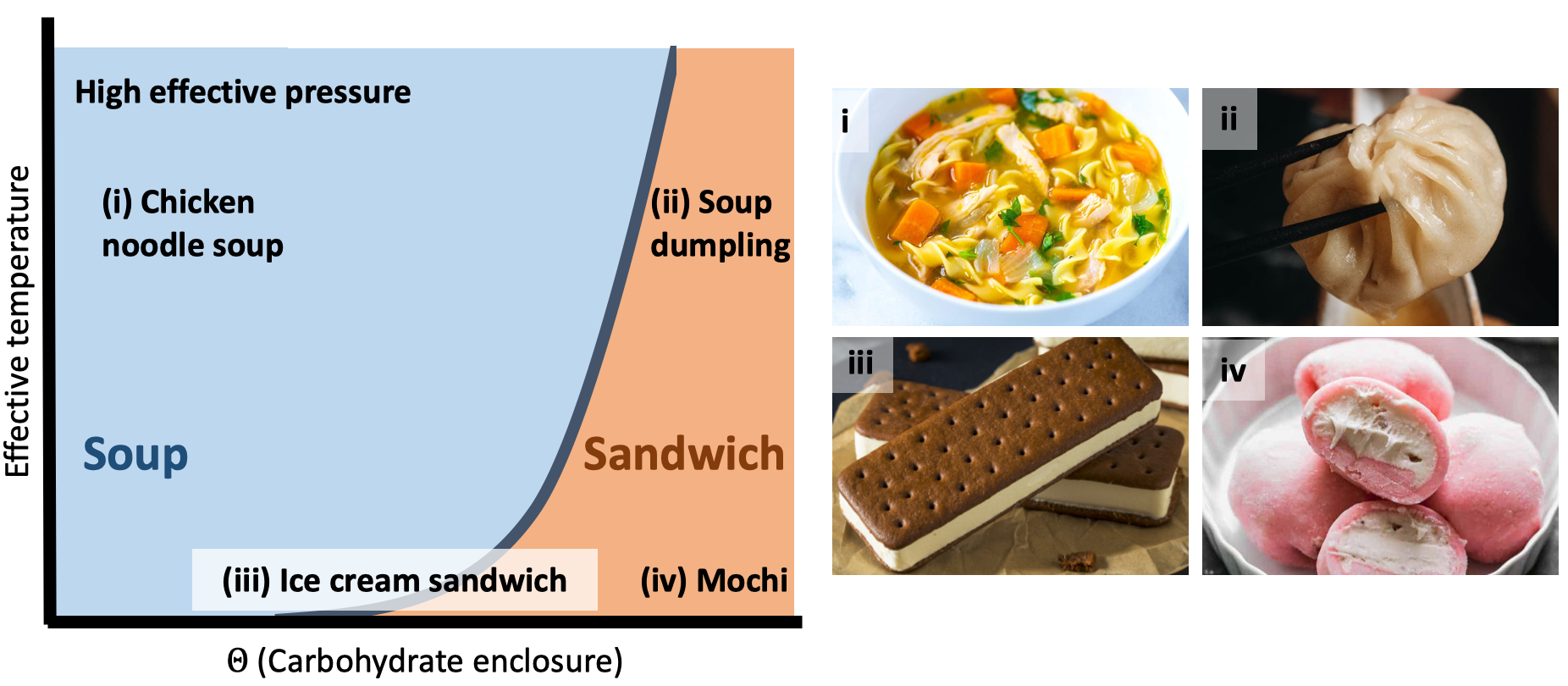}
        \subcaption{The soup-sandwich space at high effective pressure}
    \end{subfigure}
    \begin{subfigure}{0.515\textwidth}
        \centering
        \includegraphics[width=\textwidth]{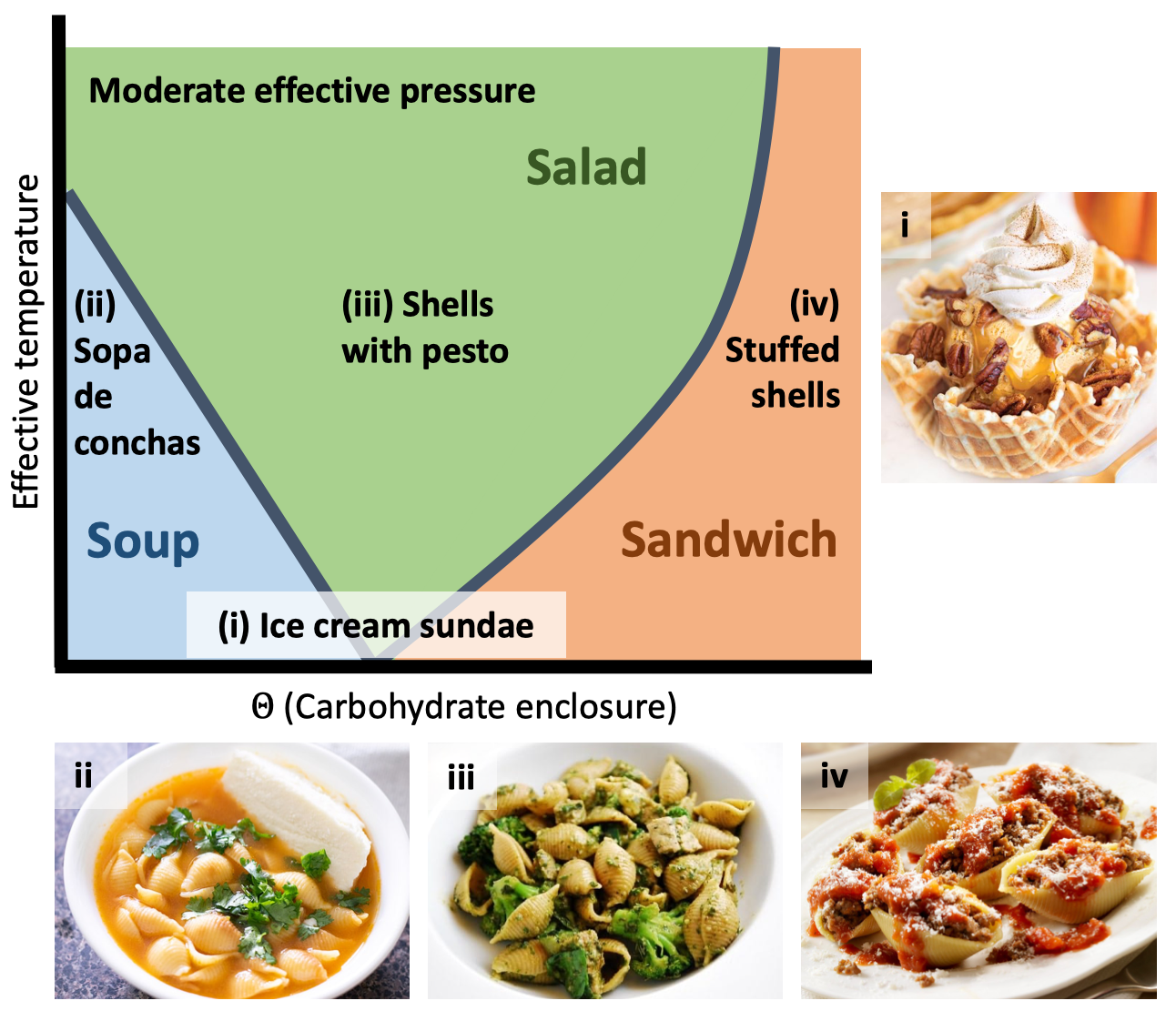}
        \subcaption{The soup-salad-sandwich space at moderate effective pressure}
    \end{subfigure}
    \begin{subfigure}{0.47\textwidth}
        \centering
        \includegraphics[width=\textwidth]{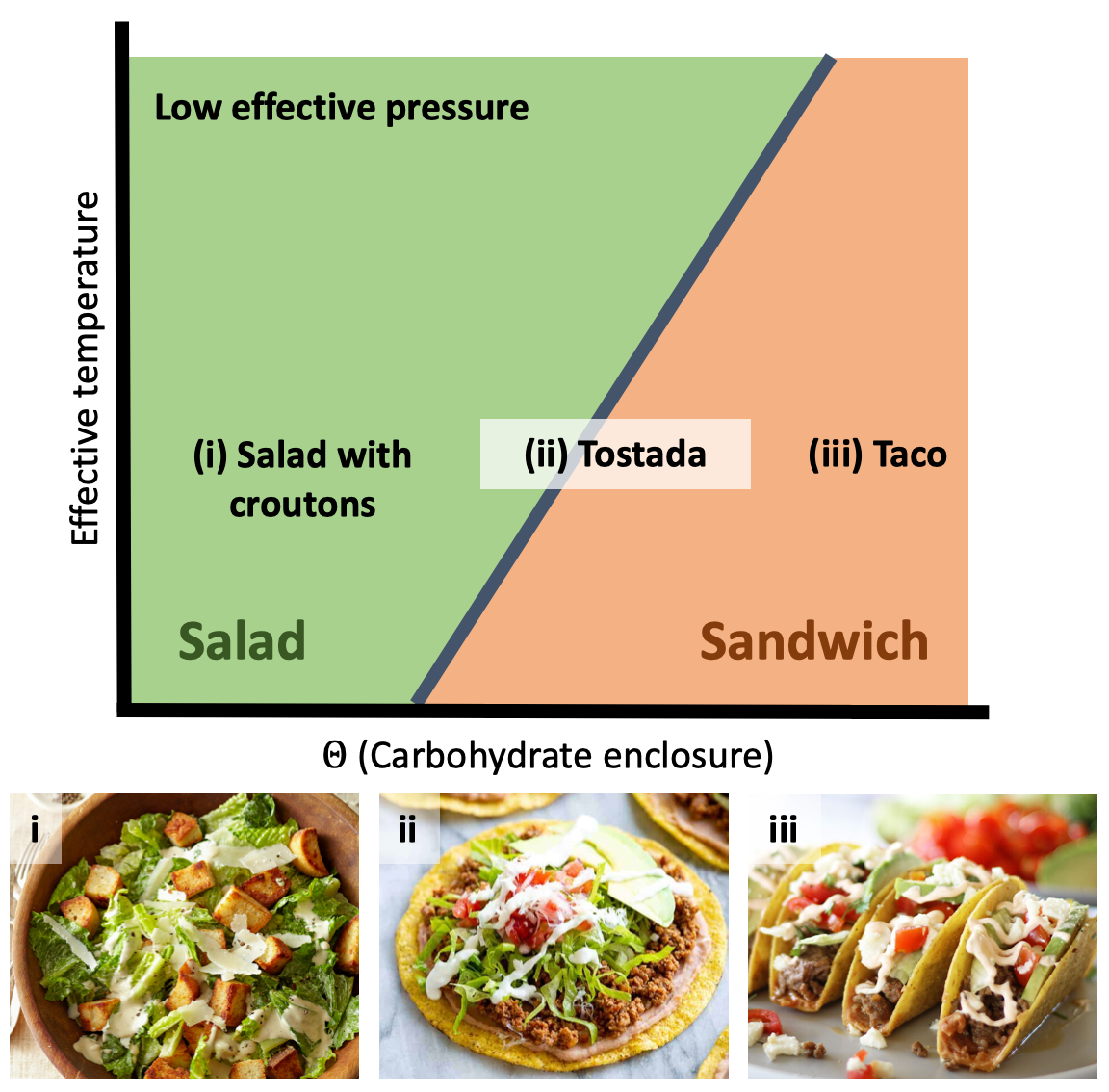}
        \subcaption{The salad-sandwich space at low effective pressure}
    \end{subfigure}
    \caption{2-dimensional slices of the full 3-dimensional phase space as a function of effective temperature and carbohydrate enclosure $\Theta$ for high, moderate, and low effective pressure. These regions of phase space reveal a rich dependence on these state variables, including a triple point of soup-salad-sandwich co-existence.}
    \label{fig:sandwich_slices}
\end{figure}

We capture these phenomena by taking slices at high, moderate, and low effective pressures and examining the sandwich transition as a function of effective temperature and carbohydrate enclosure only (See Fig. \ref{fig:sandwich_slices}). We then connect these projections to compile the full, 3-dimensional diagram of the phase space (Fig. \ref{fig:phase_diagram}). In each of the slices, the critical $\Theta_c$ line represents transitional sandwich foods. In certain regimes, the phase of the dish might be ambiguous and difficult to classify analytically. Thus, in these regions it is necessary that we to turn to an experimental probe of the system. The design of the experiment is simple; namely, the research question is: Can you pick it up by the carb and eat it? If so, the dish is a sandwich.

\subsection{Case Studies} \label{sandwich_cases}
At high effective pressure (Figure 5(a)), we find ourselves in a region of phase space with a transition between soup and sandwich phases that depends nonlinearly on the carbohydrate enclosure. At high effective temperature and low carb enclosure, we have a soups, including dishes such as chicken noodle soup; when this fluid at the same effective temperature and pressure is enclosed completely by a carbohydrate, it becomes a sandwich, namely a soup dumpling. At lower effective temperature, the necessary carbohydrate enclosure for the experimental probe of sandwichness (i.e. can you pick it up and eat it?) becomes much less due to the solid (or glassy) state of the internal components. Here, an ice cream sandwich or mochi are both solidly within the sandwich phase even at varying carbohydrate enclosure.

At moderate effective pressures (Figure 5(b)), there exists a triple point of phase co-existence between all three phases in the Triple S space. This triple point includes dishes such as an ice cream sundae in a waffle cone- a glassy soup topped with soup and salad components and contained within a carbohydrate. At lower carbohydrate enclosures, we find soups (such as sopa de conchas) that transition to a salad dish (shells with pesto) as the liquid penetrates less of the scaffold and the ingredient components are therefore embedded in a less dense matrix. Finally, at very high carbohydrate enclosures, the liquid sauce is entirely enclosed within the carb shell (stuffed shells), allowing a courageous eater to potentially pick this sandwich up and eat it.

At low effective pressure (Figure 5(c)), the behavior simplifies to a salad-sandwich transition with linear dependence on carbohydrate enclosure. For an example of a phase transition in this region, consider the constant effective temperature dishes of a salad with croutons (salad), tostada (sandwich), and taco (sandwich). This transition occurs simply due to the increasing carb enclosure, until the internal components can be fully supported by the enclosing carbohydrate.

\subsection{The Effects of Carbohydrate Component Ordering} 
\label{carb_ordering}

Upon introduction of carbohydrate enclosure, questions will naturally arise regarding continuous encasing carbs versus discrete carb elements. Particularly, one might wonder about the effects of the discrete carbs' order and symmetry on the phase of the system. It could be that there exists a yet unexplored, fourth axis of the Triple S space. This fourth axis could describe the ordering of the discrete carbs, mapping to the liquid crystal universality class of phase transitions. For example, the smectic or nematic transition could indicate a transition of the dish to a sandwich phase. A potentially compelling case study that may justify further analysis is that of spaghetti, which is solidly in the soup region when served with a lot of sauce, but is in the salad region when served only with mountains of shakey parmesan cheese. Cooked spaghetti could never fall in the sandwich phase because the carb fragments are firmly disordered or, at best, nematic (due to their lack of positional or orientational order). An additional case study may be lasagna, which sits at the soup-salad-sandwich triple point, but only because of the smectic ordering of the carb components. Baked ziti, however, has all the same components but sits on the soup-salad transition line since the noodles are disordered. While these cases all point to potentially rich areas of further study, a full exploration is beyond the scope of this work.

\section{Conclusion} 
\label{conclusion}
It is tempting to get bogged down in the microscopic status of a dish’s components when attempting to categorize the dish. However, at a certain point, we must borrow the philosophy of renormalization, and account for the fluctuations from the base, mean phase by zooming out and examining their overall, cumulative effect. At the end of the day, we must put details aside and honestly ask ourselves: “Is this a soup, a salad, or a sandwich?” Most of the time, the answer you blurt out with the least amount of thought is the most true. Concurrently we must allow for, and frankly encourage, a willingness to straddle the divides between phases and permit simultaneous truths. This is the delicacy of phase behavior: our ability to categorize the intrinsically similar while also relishing the messiness of phase co-existence, triple points, and criticality. 

Finally, to remove any of the reader's lingering uncertainty: yes, a hotdog is a sandwich.

\section{Acknowledgements}
ML and CM would both like to thank Dr. Dinah Melk, Mr. Bigg Snoup Melk, Karrot Cayke, Tayto Conqueso, S.N. Vinny, and E. Moureg for their thoughts and feedback on this project (and for tolerating innumerable heated conversations on the topic). This research was made possible by Good Soup\texttrademark, Massaged Kale Salad, and Hot Off the Panini Presses, Inc. The only competing interest to declare is our actual research desperately begging us for the time we dedicated to this paper instead. We apologize to our advisors, and promise to them that we never worked on this paper during times that we would have been otherwise productive. 

\bibliographystyle{unsrt}
\bibliography{phase_transitions}

\end{document}